\documentstyle[12pt]{article}

\newcommand{\be}{\begin{equation}}
\newcommand{\ee}{\end{equation}}
\newcommand{\bea}{\begin{eqnarray}}
\newcommand{\eea}{\end{eqnarray}}

\begin{document}
\begin{titlepage}


\vspace{1in}

\begin{center}
\Large
{\bf Initial Conditions in String Cosmology}

\vspace{1in}

\normalsize
 
\large{Dominic Clancy$^{1a}$, James E. Lidsey$^{2b}$ 
\& Reza Tavakol$^{1c}$}
 
\normalsize
\vspace{.7in}
 
$^1${\em Astronomy Unit, School of Mathematical Sciences,  \\
Queen Mary \& Westfield College, Mile End Road, LONDON, E1 4NS, U.K.}

\vspace{.2in}
$^2${\em Astronomy Centre and Centre for Theoretical Physics, \\
University of Sussex, BRIGHTON, BN1 9QJ, U.K.}

\end{center}

\vspace{1in}

\baselineskip=24pt
\begin{abstract} 
\noindent
We take a critical look at a recent conjecture concerning the past 
attractor in the pre-big-bang scenario. We argue that 
the Milne universe is unlikely to be a general past attractor for such 
models and support this with a number of examples. 
\end{abstract}

PACS NUMBERS: 98.80.Cq

\vspace{.7in}
$^a$Electronic address: dominic@maths.qmw.ac.uk
 
$^b$Electronic address: jel@astr.cpes.susx.ac.uk
 
$^c$Electronic address: reza@maths.qmw.ac.uk
 
\end{titlepage}


An important issue in cosmology, whether in the context of
general relativity or string theory, 
is that of ``naturalness''. This  in turn
translates into the question of
initial conditions. One way of studying this
question is to ask whether there exists an
``attractor'' whose basin of attraction  
(i.e. the set of initial conditions which evolve 
to this state) has a large or full measure
in the space of all possible initial data.

In its fullness, this is an impossible
question to address at present 
for a number of reasons. Despite dramatic recent progress 
in our understanding of M-theory, 
we still lack a definitive non--perturbative formulation of string theory.
(For a recent review, see, e.g., Ref.~\cite{d}).  
This raises the question of whether the 
results obtained within the context of the perturbative 
effective actions are representative and robust with 
respect to higher-derivative and loop corrections. 
Furthermore, the nature of the generic attractor is not 
known even at this reduced level because the resulting equations 
are non-linear partial differential equations. 
Consequently, additional restrictions such as spatial homogeneity 
must be imposed to drastically reduce the 
complexity of these equations.
The crucial point here is to establish those attractors that 
have the largest basins and are therefore the most `natural' 
at this level of approximation. 

An interesting recent development within string cosmology has
been the {\em pre-big-bang} scenario~\cite{pbb}. In this picture, 
an accelerated, inflationary expansion is driven by the 
rapid increase of the gravitational (string) coupling, 
as parametrized by the dilaton field. 
The fundamental postulate of the pre-big-bang cosmology is that 
the initial data for inflation lies well within the 
perturbative regime, where the curvature and coupling are very 
small. Inflation then proceeds 
for sufficiently homogeneous initial conditions, 
where time derivatives are dominant with respect to
spatial gradients, and the universe evolves into a high curvature and 
strongly-coupled regime. At the level of the 
lowest-order effective action, the final state of this evolution 
is singular in both the curvature and coupling, but it has been 
proposed that higher-order corrections should become important 
at the string scale~\cite{ho}. 

The behaviour of 
pre-big-bang cosmology in the asymptotic 
past before the onset of inflation has also 
been addressed~\cite{early,early1,early2}. 
It was recently conjectured that pre-big-bang 
inflation generically evolves out of an initial state
that 
approaches the Milne universe in the semi-infinite past,  
$t \rightarrow -\infty$, where $t$ represents synchronous time~\cite{early1}. 
The Milne universe may be mapped onto the future (or past) light cone of the 
origin of Minkowski spacetime and therefore corresponds to a 
non-standard representation of the string 
perturbative vacuum. It is flat spacetime expressed in an expanding frame: 
\be
\label{m5}
ds^2 =-dt^2 +t^2 \left( dx^2 + e^{-2x} (dy^2 +dz^2 ) \right)  .
\ee
The proposal is that the Milne 
background represents an early time attractor, with a large
measure in the space of initial data~\cite{early1}.
If so, this 
would provide strong justification for the postulate that 
inflation begins in the weak coupling and curvature regimes
and would render the pre-big-bang assumptions regarding the 
initial states as `natural'.

The aim of this paper is to take a careful look at this conjecture.
Since pre-big bang inflation must begin in the perturbative regime, 
the dynamics of the universe 
is well approximated by the string effective action. To 
lowest-order in the inverse string tension, the massless 
bosonic excitations in the 
Neveu-Schwarz/Neveu-Schwarz sector of the theory are the dilaton, 
$\Phi$, the 
graviton, $g_{\mu\nu}$, 
and the anti-symmetric two-form potential, $B_{\mu\nu}$, with 
field strength, $H_{\mu\nu\lambda} \equiv 
\partial_{[\mu}B_{\nu\lambda ]}$. The effective action is \cite{string}
\be
\label{effact}
S=\int d^4 x \sqrt{-g} e^{-\Phi} \left[ R + \left( 
\nabla \Phi \right)^2 -\frac{1}{12} H_{\mu\nu\lambda}H^{\mu\nu\lambda} 
\right]  ,
\ee
where $R$ is the Ricci curvature scalar and $g \equiv {\rm det}g_{\mu\nu}$.

Fundamental strings sweep geodesic surfaces 
with respect to the string frame metric, $g_{\mu\nu}$,
but it is convenient to discuss the dynamics 
in the conformally related Einstein frame, where the 
dilaton is minimally coupled to the graviton: 
\be
\tilde{g}_{\mu\nu} \equiv e^{-\Phi} g_{\mu\nu}.
\ee
Moreover, the  field strength 
of the two-form potential is dual to a one-form 
in four dimensions, i.e., $H^{\alpha\beta\gamma} \equiv 
e^{\Phi} \epsilon^{\alpha\beta\gamma\delta} \nabla_{\delta} \sigma$, 
where $\epsilon^{\alpha\beta\gamma\delta}$ is the covariantly constant 
four-form. This one-form  may then be interpreted as the field 
strength of a pseudo-scalar axion field, $\sigma$. 
The effective action (\ref{effact}) is therefore 
equivalent to
\be
S=\int d^4 x \sqrt{-\tilde{g}} \left[ 
\tilde{R} -\frac{1}{2} \left( \tilde{\nabla} \Phi \right)^2 
-\frac{1}{2} e^{2\Phi} \left( \tilde{\nabla} \sigma \right)^2 
\right]  .
\ee
For the class of models we consider, 
a massless, minimally coupled scalar field $\phi$ (where 
the gradient of the scalar field
is a timelike vector) may be interpreted
in terms of a stiff perfect fluid:
\bea
p=\rho  &=& \frac{1}{2} \phi_{,\alpha} \phi^{, \alpha} \\
u_{\alpha} &=& \frac{\phi_{,\alpha}}{\left( \phi_{,\beta} \phi^{, \beta}
\right)^{1/2}}   ,
\eea
where $p$ is the pressure, $\rho$ is the energy density and
$u_{\alpha}$ is the fluid four-velocity.
Since both the dilaton 
and axion fields are massless, the energy-momentum 
tensor in the Einstein frame is then 
equivalent to that of a stiff perfect fluid, 
with the equation of state $p=\rho$.

In the synchronous frame, where $g_{00} =-1$ 
and $g_{0i} =0$, the spacetime metric may be written in the form
$ds^2 =-dt^2 +h_{ij}dx^idx^j$ $(i,j =1,2,3)$. The Einstein field 
equations in this frame are then given by~\cite{ll}
\bea 
\label{efe1}
{R^0}_0 &=& -\frac{1}{2} \frac{\partial}{\partial t} {\chi^i}_i
-\frac{1}{4} {\chi^j}_i {\chi^i}_j = {T^0}_0 -\frac{1}{2} T \\
\label{efe2}
{R^0}_i &=& \frac{1}{2} \left ({\chi^j}_{i ; j}
-{\chi^j}_{j ; i} \right ) = {T^0}_i\\
\label{efe3}
{R^j}_i &=& - ~{^{(3)}}{R^j}_i - \frac{1}{2\sqrt{h}}
\frac{\partial}{\partial t} \left ( \sqrt{h} {\chi^j}_i \right )
= {T^j}_i -\frac{1}{2} {\delta^j}_i T   ,
\eea
where $\chi_{ij} \equiv \partial h_{ij} / \partial t$, 
${^{(3)}}{R^i}_j$ is the 
Ricci curvature tensor constructed from the three-metric $h_{ij}$, 
$h \equiv {\rm det} h_{ij}$, a semi-colon 
denotes covariant differentiation with respect to $h_{ij}$
and  units are chosen such that $8\pi G =1$. 

We begin by briefly reviewing the argument that
the only non-singular fixed point of the 
Einstein field equations (\ref{efe1})-(\ref{efe3})
with a massless scalar field is flat space \cite{early,early1}.
It can be shown that the fixed 
points exist 
either at infinity or at $\dot{\phi} =\chi_{ij} =\nabla^i \nabla_i \phi =0$. 
These two possibilities correspond to a singularity or a stationary  
field, respectively. The latter further 
implies that $\nabla_i \phi =0$ if boundary terms 
are neglected and it then follows from the field 
equations that ${^{(3)}}R_{ij} =0$. 
Since in three dimensions a vanishing Ricci tensor implies 
a vanishing Riemann tensor, it is concluded that 
the only non-singular 
past attractor corresponds to flat spacetime. Buonanno {\em et al.}~\cite{early1} then conjecture that those pre-big-bang 
models that are negatively-curved 
and sufficiently isotropic generically evolve out of the Milne state. 

However, we 
recall that an important feature of synchronous reference 
frames is that they are manifestly {\em not} stationary, 
as emphasized by Landau and Lifshitz~\cite{ll}. In other words,  
a gravitational 
field can not be constant in such a frame (i.e. of the form
$\chi_{ij}=0$) while possessing a non-zero energy-momentum tensor,
as can be seen directly from Eq. (\ref{efe1}). Thus, 
synchronous frameworks are not compatible with non-vacuum,
stationary fields (fixed points).  
As a result, the Milne universe as a seemingly sole past attractor
arises as a direct consequence of working in the synchronous frame, 
because in this frame the only non-singular stationary solution (fixed point)
is necessarily flat space. 

Now,  given that generically one expects to have 
a non-zero energy-momentum tensor at the onset of inflation, 
a more natural question is to ask 
whether the Milne universe is likely to be the past attractor 
with the largest measure of initial states if one starts with a 
non-vacuum, pre-big-bang universe 
and runs it backwards towards $t \rightarrow -\infty$. 
In the following 
we shall argue that this is unlikely. We should emphasize here
that the Milne universe may still  be an attractor for a
certain class of cosmologies. The question we 
address is that of its likelihood
(i.e. the size of the corresponding basin and 
whether or not it carries a full measure). 

To substantiate the above discussion,
we proceed to present some concrete examples for which the 
past attractor is {\em not} the Milne universe. 
We consider families of models that are often considered in
theoretical cosmology, and ask what is the generic initial state among
these particular sets of models out of 
which an inflating pre-big-bang cosmology evolved. 

Without loss of generality, 
we may work in the Einstein frame, since we are 
interested in asymptotic states where the dilaton and axion fields approach 
constant values. In this case, the string and Einstein frames 
become equivalent. It then follows from the time 
symmetry of the Einstein field equations that we may 
gain insight into the nature of the past attractor of 
pre-big-bang cosmology by instead considering the future asymptotic 
states of classical cosmological solutions to Einstein gravity containing 
a stiff perfect fluid. 

We begin with the spatially homogeneous Bianchi models. (For a review of 
the Bianchi classification scheme, see, e.g., Ref.~\cite{ryan}). 
These universes 
admit a three-dimensional Lie group of isometries acting simply 
transitively on the spacelike hypersurfaces, $t={\rm constant}$. 
In the Ellis-MacCallum classification, the models are separated
into two classes, A and B, depending on the specific
group type~\cite{em}. Bianchi models are
referred to as orthogonal cosmologies when the fluid flow is
orthogonal to the group orbits. Otherwise they represent
`tilted' models. 
Tilted models are homogeneous to observers with
world lines directed orthogonally to the $t={\rm constant}$ hypersurfaces,
but appear to be inhomogeneous to those observers that comove with
the fluid.

It is natural to suppose that 
the ability of a given model to approach the vacuum Milne state at 
late times should be related to the question of whether it 
is able to isotropize. Collins and Hawking~\cite{ch}
have proved a theorem stating 
that modulo some very general assumptions about the matter 
fields (that are satisfied by an orthogonal  stiff fluid), 
{\em the set of spatially 
homogeneous models that isotropize at late times is of measure zero
in the space of homogeneous initial data}. Indeed, 
they prove that the only Bianchi types that could 
possibly isotropize at arbitrarily late times are the types 
I, V, ${\rm VII}_0$ and ${\rm VII}_h$. It turns out that types 
I, V and ${\rm VII}_0$ can approach isotropy. 
However, the type ${\rm VII}_h$, 
which is the only model amongst this subgroup that 
has non-zero measure in the
space of all homogeneous models, in general does not.
In fact, a necessary condition 
for a type ${\rm VII}_h$ model to isotropize is that it must tend to a $k=-1$ 
Friedmann-Robertson-Walker (FRW) universe. Even though 
this model does approach Milne 
asymptotically, it is nevertheless of measure zero
in the set of ${\rm VII}_h$ models.
This suggests that the 
set of type ${\rm VII}_h$ pre-big-bang models driven by a dilaton 
field that would have 
evolved out of the Milne state is of measure zero in the space 
of type ${\rm VII}_h$ cosmologies.
 
We now consider the class B models in more detail. 
This class contains the Bianchi types IV, V, ${\rm VI}_h$ and 
${\rm VII}_h$. It is known that all 
orthogonal class B perfect fluid models expand indefinitely 
into the future $(t>0)$. Hewitt and Wainwright  
have proved an important theorem for these models by employing 
a dynamical systems approach to homogeneous cosmology 
\cite{hw}. They have shown that 
{\em all orthogonal class B Bianchi models with a stiff perfect fluid, 
apart from a set of measure zero,  
are asymptotic in the future to a plane wave state and asymptotic 
in the past to the Jacobs Bianchi I solution}. We may interpret this 
result in the pre-big-bang context by interchanging the past 
and future states. The Jacobs stiff perfect fluid solution then 
describes the evolution of the universe towards the singularity 
\cite{jac}. It contains 
two essential parameters and may be interpreted as the generalization 
of the vacuum Kasner solution to include a dilaton field 
\cite{kas}. It is precisely this 
solution that corresponds to dilaton-driven, pre-big-bang inflation 
in the string frame when the parameters satisfy appropriate conditions~\cite{dominic}.  
In general, however,  the initial state does not correspond 
to a region of flat space, but rather to a plane wave. Two exceptions are 
the orthogonal types V and ${\rm VI}_{-1}$. (The latter model is sometimes 
referred to as the type III). In these cases, the models do indeed 
approach flat space as we trace their behaviour back to 
$t \rightarrow -\infty$. 

Further insight may be gained by 
considering the type ${\rm VII}_h$ model.
In an appropriate coordinate frame, the most 
general type ${\rm VII}_h$ metric may be written in the 
form \cite{s}
\be
ds^2 =a^{-2} \left( -d \tau^2 +dx^2 \right) +e^{2(\lambda -x)} dS^2  ,
\ee
where
\bea
dS^2 &=& \cosh \mu \left( dy^2 +dz^2 \right)
\\ &\,&
-\sinh \mu \left[ \left(  dz^2 -dy^2  \right) \cos 2(kx +\varphi ) +
2 dydz \sin 2(kx +\varphi ) \right]\nonumber,
\eea
the four variables  $\{ a , \mu, \lambda , \varphi \}$ are functions 
only of $t$, $k^2 \equiv h^{-1}$ 
and we have defined a time parameter $\tau \equiv \int^t dt_1 a(t_1)$. 
The Einstein field equations (\ref{efe1})--(\ref{efe3}) 
containing an orthogonal perfect fluid are given by~\cite{bs}
\be
\label{71}
\lambda^{\prime \prime} 
+2\lambda^{\prime 2} -2 -\frac{1}{2} (\rho -p )= 0,
\ee
\be
\label{72}
\nu^{\prime} 
+2 \lambda^{\prime} \nu +2\sinh 2\mu \left( k^2 - \varphi^{\prime 2} 
\right) =0, 
\ee
\be
\label{73}
\left( \varphi^{\prime}  \sinh^2 \mu \right)^{\prime} +2 
\lambda^{\prime} \left( \varphi^{\prime} \sinh^2 \mu \right) +2k 
\sinh^2 \mu =0,
\ee
\be
\label{74}
\alpha^{\prime} +2\alpha \lambda^{\prime} +2\left( 1+k^2 \sinh^2 \mu \right) 
+\frac{1}{2} (\rho -p )= 0, 
\ee
\be
\label{75}
\alpha +\lambda^{\prime} -k \varphi^{\prime} \sinh^2 \mu = 0,
\ee
\be
\label{76}
\frac{1}{4} \nu^2 +\left( 3+ k^2 \sinh^2 \mu \right) \left( 
1- \lambda^{\prime 2} \right) +\sinh^2 \mu \left( k \lambda^{\prime} 
+ \varphi^{\prime} \right)^2 + \rho =0,
\ee
where $\nu \equiv \mu^{\prime}$, $\alpha \equiv a^{\prime}/a$, 
a prime denotes differentiation 
with respect to $\tau$ and the energy density and pressure have been 
scaled by a factor $a^{-2}$. 

These equations for a stiff perfect fluid 
are {\em identical} to those of the vacuum model, with the exception 
of Eq. (\ref{76}), where there is an additional term due to 
the energy density. 
This allows the asymptotic form of the general type 
${\rm VII}_h$ stiff perfect fluid solution to be established~\cite{s,bs}. 
Eq. (\ref{71}) can be solved in full generality 
and the solution approaches $\lambda^{\prime} \rightarrow 1$ as $\tau 
\rightarrow \infty$. Thus, the second term on the left-hand side of 
Eq. (\ref{76}) vanishes, but since the remaining terms in this 
equation are all 
positive definite, they must  each  vanish separately. It follows, therefore, 
that $\rho \rightarrow 0$, 
$\varphi^{\prime} \rightarrow -k$, $\mu 
\rightarrow \mu_0 ={\rm constant}$ and $\alpha \rightarrow 
(1+k^2 \sinh^2 \mu )$. This metric  
corresponds to the vacuum type ${\rm VII}_h$ plane wave solution
first derived by Doroshkevich, Lukash and Novikov~\cite{dln}. 
It may be interpreted as representing two 
monochromatic, circularly-polarized 
gravitational waves traveling with 
constant amplitudes in opposite directions along the 
$x$-axis~\cite{lukash}. 
This plane wave solution would have to isotropize if it were to approach 
the Milne state at large $t$, but 
it is known that this does not occur. 
Thus, all but a set of measure zero orthogonal 
stiff perfect fluid type ${\rm VII}_h$ 
cosmologies asymptotically approach 
a plane wave and can not be written in the Milne form. 
An example of a 
type ${\rm VII}_h$ solution that does tend to Milne is the 
particular $h=4/11$ solution due to Barrow~\cite{barrow}. 

It is important to note that the orthogonal models are special. It 
is therefore pertinent to consider the effect 
that tilt has on the asymptotic behaviour of the 
class B models. This is interesting because 
tilt can be viewed as a form of inhomogeneity. 
Of particular interest is the type V model, 
since the orthogonal stiff fluid solution does approach 
the Milne model \cite{hwprd}. We consider the tilted type V
stiff fluid solution found by Maartens and Nel~\cite{mn}: 
\bea 
ds^2 &=& e^{2k} \left( -d\tau^2 +dx^2 \right) + 
r \left( f dy^2 +f^{-1} dz^2 \right) \\
\rho &=& e^{-2k} \left( {\sigma_{,\tau}}^2 - 
{\sigma_{,x}}^2 \right)   ,
\eea
where a comma denotes partial differentiation, the 
functions $\{ r, f, k, \sigma \}$ are defined by  
\bea
r &=& e^{-2qx} \sinh 2q \tau, \nonumber \\
f &=& \left( {\rm tanh}  q \tau \right)^m, \nonumber \\
e^{2k} &=& \left(  \sinh 2q \tau  \right)^{\alpha^2 + \beta^2 +(m^2-1)/2}
\left( {\rm tanh} q \tau \right)^{2 \alpha \beta} \nonumber, \\
e^{\sigma} &=& \left( {\rm tanh} q \tau  \right)^{-\alpha} 
\left( \sinh 2q \tau  \right)^{-\beta} e^{2\beta q x}, 
\eea
and the constants $\{ m, \alpha , \beta \}$ satisfy the constraint
\be
m^2 -3 +2\left( \alpha^2 -\beta^2 \right) =0.
\ee
The scalar function $\sigma$ may be 
interpreted as a minimally coupled scalar field. 
The solution is tilted if $\beta \ne 0$ and the fluid flow is 
orthogonal to the surfaces of homogeneity 
if $\beta =0$. 

In the asymptotic limit $\tau  \rightarrow 
\infty$ $(t \rightarrow \infty )$, the metric reduces to 
\be
\label{tiltv}
ds^2 =e^{2 (2\beta^2+1) \tau} \left( -d \tau^2 +dx^2 \right) 
+ e^{2 \tau -2x} \left( dy^2 +dz^2 \right),
\ee
where we have specified $q=1$ without loss of generality. 
The matter lines become null and this corresponds to the case of 
extreme tilt. Eq. (\ref{tiltv})  
is a locally rotationally symmetric (LRS) model and can be 
identified with the equilibrium point labeled ${\cal{H}}$ 
in the notation of Hewitt and Wainwright~\cite{hwprd}.
It can be shown that for this model the dimensionless variable defined 
by $\Sigma_+ \equiv \sigma_+/\theta$, where 
$\sigma_+$ is a shear parameter and $\theta$ is the rate of expansion 
scalar, tends to a non-vanishing constant given by  
$\Sigma_+ =-2\beta^2/(3+2\beta^2)$.  

Moreover, if we define the light-cone coordinates: 
\bea
u &\equiv & \exp  \left[ (1+2\beta^2)(\tau -x) \right], \nonumber \\
v &\equiv &\frac{1}{(1+2\beta^2)^2} \exp \left[ 
(1+2\beta^2 )( \tau +x) \right]   .
\eea
Eq. (\ref{tiltv}) may be written in the form
\be
ds^2 =-dudv +u^{2/(1+2\beta^2)} \left( dy^2 
+dz^2 \right).
\ee
This demonstrates that in general, Eq. (\ref{tiltv})  
corresponds to a conformally flat, homogeneous plane wave. 
It reduces to the Milne form of flat space only 
in the orthogonal model, where $\beta =0$, and only in this special case does 
the dimensionless shear parameter, $\Sigma_+$, vanish. 

It is also important to consider the effects that spatial 
inhomogeneity may have on the possible set of initial 
conditions. The simplest class of 
inhomogeneous models are those that break homogeneity in only 
one direction $(x)$. In general, these models admit an abelian  
group of isometries, $G_2$, with orbits corresponding to 
space-like two-surfaces. (For a review, see, e.g., Ref.~\cite{ccm}). 
In the case where the $G_2$ admits 
two hypersurface-orthogonal Killing vector fields,
the line-element may be written in the diagonal form~\cite{ccm}: 
\be
ds^2 =e^f \left( -d\xi^2 +dx^2 \right) +R 
\left( e^pdy^2 +e^{-p} dz^2 \right)   ,
\ee
where $f=f(\xi , x)$ and $p=p( \xi , x )$ represent the longitudinal 
and transverse parts of the gravitational field, respectively. 
The gradient of the function $R=R(\xi ,x)$ determines the 
properties of the model and the 
Killing vectors are $\partial /\partial y$ and $\partial / \partial z$. 

As an example, we 
consider the class of spatially compact cosmologies with a three--torus 
topology $S^1 \times S^1 \times S^1$. In this case, one may 
specify $R=\xi$ without loss of generality. The general 
solution to the Einstein-scalar field 
equations for these models has been found previously by 
Charach and Malin~\cite{cm}, who also discussed 
the asymptotic behaviour of the solution 
in the late time (high-frequency) limit, $\xi \rightarrow \infty$. 
In general, the scalar field, $\phi$, and transverse mode, 
$p$, decouple and both satisfy the one-dimensional
cylindrical wave equation. In particular, the transverse 
part of the metric is given by~\cite{cm}
\be
p=p_0 +\alpha_0 \ln \xi +\sum_{n=1}^{\infty} \cos 
\left[ n (x -x_n ) \right] \left[ 
A_n J_0 (n\xi) +B_n N_0 (n\xi) \right]   ,
\ee
where $\{ p_0 ,\alpha_0 ,A_n, B_n , x_n \}$ are arbitrary constants and $\{ 
J_0 , N_0 \}$ are Bessel and Neumann functions of order zero. Formally, 
a similar expression may be written for the scalar field and 
the general form of the longitudinal mode, $f$, can then be determined. 

It was found that the model asymptotically 
evolves into the Doroshkevich, Zeldovich and Novikov 
(DZN) universe~\cite{dzn}. This 
corresponds to an anisotropic, spatially homogeneous background 
with a null fluid  composed of collisionless flows of massless scalar 
particles and gravitons. The late-time limit 
of the line-element is given in synchronous coordinates by 
\be
\label{lateI}
ds^2 =-dt^2 +a_1^2(t) dx^2 + a_2^2(t) dy^2 +a_3^2(t) dz^2   ,
\ee
where
\be
a_1 \propto t , \qquad a_2 \propto (\ln t )^{(\alpha_0 +1)/2} , 
\qquad a_3 \propto (\ln t)^{(1-\alpha_0 )/2}
\ee
and Eq. (\ref{lateI}) does not reduce to the Taub form 
of flat space, $ds^2 =-dt^2 +t^2 dx^2 + dy^2 +dz^2$.  

In conclusion, we have argued that 
the recent conjecture that the Milne universe is a past 
asymptotic attractor for pre-big-bang cosmologies 
is a consequence of studying the dynamics in the 
synchronous frame. To substantiate this, we have 
provided a number of examples for which the past attractor is not 
the Milne form of flat space as given by Eq. (\ref{m5}). This
included the class of orthogonal, anisotropic Bianchi B models. 
This class has a non-zero measure in the space of homogeneous initial data.  
A plane wave background is the attractor with a full
measure of initial states for this class. 
Even in the Bianchi V case, where the orthogonal 
solution does tend towards Milne, the inclusion of tilt in the fluid flow
removes this possibility. In addition, we have discussed the $G_2$ 
inhomogeneous generalization of the Bianchi I model for which the attractor
is the homogeneous DZN cosmology. 

These examples demonstrate that 
within the classes of models we have considered, 
the Milne universe is an unlikely 
past attractor for the pre-big-bang scenario. 
It should be emphasized that this does
not indicate that such a state is not allowed 
and, indeed, it has a number of attractive features. 
However, we have provided quantitative arguments that suggest 
it is not generic. Moreover, our conclusions 
clearly do not constitute a general result within all 
possible models. In view of this, it 
would be interesting to consider other classes of inhomogeneous models to 
establish the extent of this likelihood in more general settings. Work 
in this direction is in progress. 

Finally, in the above analysis 
of the orthogonal Bianchi models, we assumed that the 
axion field was homogeneous. 
Alternatively,  one may consider the case where the two-form potential
is homogeneous, which then implies an inhomogeneous
axion field. Recently, Barrow and Kunze considered 
the most general form of the anti-symmetric field strength 
compatible with spatial homogeneity in this latter case \cite{bk}. 
They demonstrated that the
Bianchi class B types III and ${\rm VI}_{h=0,-1/2,-2}$ are the most general. 
This is in marked contrast to the case where the axion field is homogeneous, 
where the types ${\rm VI}_h$, ${\rm VII}_h$, VIII and IX have the highest 
dimension. This would seem to indicate that assuming a time-dependent 
two-form potential could lead to different conclusions
regarding the generality of the past attractor for spatially homogeneous 
pre-big-bang cosmologies. It would be interesting to investigate this 
possibility further.  

\vspace{.3in}

\centerline{\bf Acknowledgments}

\vspace{.3in}

D. C. and J. E. L. are supported by the Particle Physics and Astronomy 
Research Council (PPARC), U. K. and  
R. T. benefited from PPARC UK Grant No. L39094. We thank 
A. Feinstein, J. Maharana, G. Veneziano and J. Wainwright, 
for helpful communications.

\end{document}